# Klapdor's claim for the observation of the neutrinoless 2β-decay of $^{76}$Ge. Analysis and corrections.

Kirpichnikov I.V.
Institute for Theoretical and Experimental Physics, B.Cheremushkinskaya 25, 117218 Moscow

Heidelberg–Moscow Collaboration experimental data [1-4] with the claim for the observation of the neutrinoless 2β-decay of $^{76}$Ge has been carefully analyzed. The analysis gave an evidence that the observed 2039 kev line in the full spectrum was produced by overlapping of three unresolved peaks: ~2035.5 kev, ~2037.5 kev and ~2039.5 kev . It was indicated that the 2035.5 kev and 2039.5 kev peaks were produced by the double-coincidences of gamma-quanta in the detectors. Existence of the two background gamma-peaks at ~2035.5 kev and ~2039.5 kev was confirmed by an analysis of the background measurements with low-background Ge detectors with an active shielding [5-7]. The 2037.5 kev line could be the expected 2β0ν-decay transition of $^{76}$Ge according to the results of the improved pulse-shape analysis. Position of the line, 2037.56±0.56(stat)±1.2(syst) kev [4], was slightly below the expected, $E_{2\beta}$=2039.0±0.05 kev [8], but still within a limit for poor statistics. The most accurate value of its intensity was extracted from the 51.39 kg·y data with PSA: number of events 12.4±3.7 and $T_{1/2}$=1.98·10$^{25}$ years [2,3].

The search for the neutrinoless double beta-decay is considered as one of the most important problems of particle physics and cosmology. The process is possible if the neutrino is a massive Majorana particle. The results of the oscillation experiments [9] proved the neutrino being a massive particle. Still, the nature of the neutrino, an absolute scale of the mass and an hierarchy of the masses are unknown. The investigation of the 2β0ν-decay can solve these problems. In particular, the observation of the 2β0ν-decay is the most efficient method for solving the problem whether the neutrino is a Dirac or a Majorana particle.

The first and the only claim for the observation of the neutrinoless process (2β0ν-decay of $^{76}$Ge) was published in 2001 by a group of authors from the Heidelberg–Moscow Collaboration [1]. A peak at ~2038 kev, close to the energy of the 2β0ν-decay of $^{76}$Ge, $E = (2039.0 \pm 0.05)$ keV, was found in the background spectra of Ge detectors at a statistical significance of about 3σ. The CL of the peak was increased up to 4.2σ in the subsequent publications of the same group of authors by using the full statistics (fig.1) collected over the period between 1990 and 2003 years [2,3].

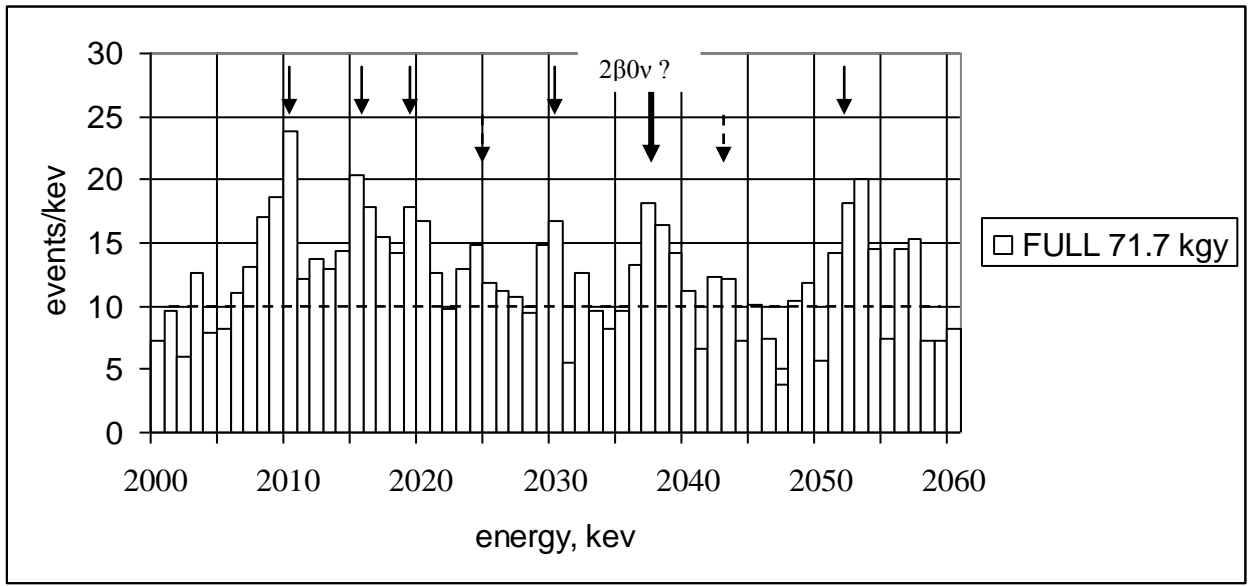

Fig.1. Full spectrum with statistics 71.7 kg·y [2,3]. The solid arrows showed 6 peaks included in the analysis . The dashed ones presented peaks to be added. The dashed straight line shows the level of the slow varying (constant) component (C≈10 events/kev). An overestimation of this component was indicated by the best fit procedure including eight peaks (C≈8 events/kev).

The final position of the peak was found E=(2038.07 ±0.44) kev. On the basis of the analysis of possible radioactive background sources, the authors claimed that the observed effect should be associated with the process of neutrinoless double-beta decay. The peak intensity corresponded to the most probable value $T_{1/2}(0\nu) = 1.2 \cdot 10^{25}$ years.

However the interpretation of the peak as a signature of the 2β0ν-decay of $^{76}$Ge was dubious. The authors of [2,3] included into their analysis 6 levels within an energy interval (2000-2060)kev. It was indicated that one should include in the analysis two extra peaks at least, at 2023 kev and at 2044 kev [4,5]. The nature of several peaks has not been understood. The probability to find an unknown background peak in the close vicinity of the expected $^{76}$Ge 2β0ν-decay line seemed rather high.

The additional information concerning a nature of the 2038 kev peak was necessary. Such information could be provided by the analysis of the shapes of the signals – the Pulse Shape Analysis (PSA). The authors of the experiment have fully understood it. So the large bulk of H-M statistic (51.39 kg·y of the total 71.7 kg·y) was collected with the registration of the shapes of the signals. Several algorithms of the data treatment were proposed with the two different PSA methods [10,11], which provided the extraction from the full spectrum the so called SSE spectrum (Single Site Events; all signals of the 2β0ν-decay should be of this type, fig.2). The background index in the (2000 -2060) kev interval was reduced three times, from 0.158 events/kev·kgy in the FULL spectrum, down to 0.054 events/kev·kgy in the SSE spectrum.

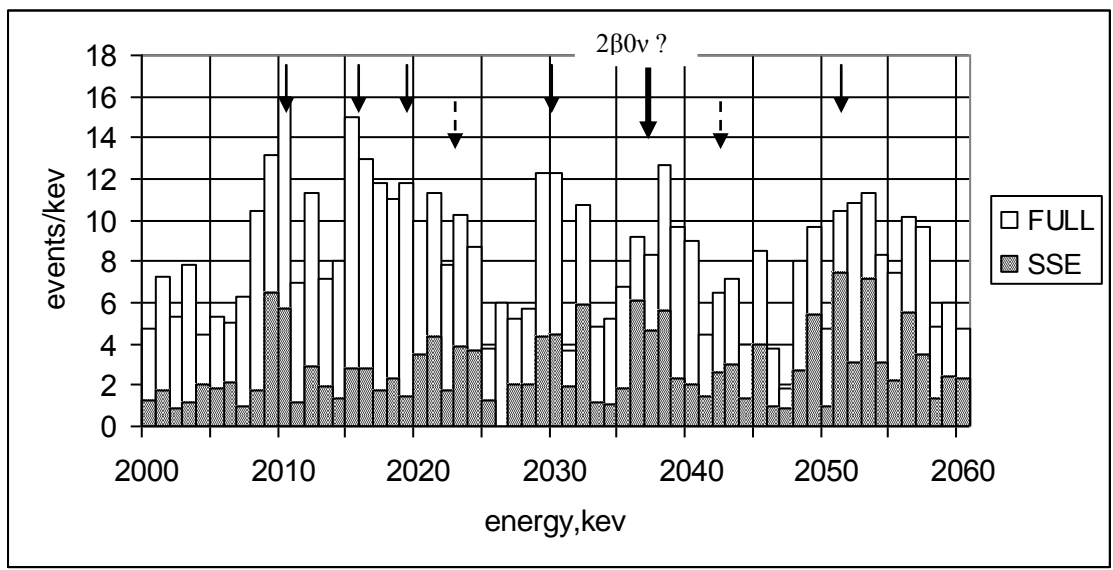

Fig.2. Results of the the PSA: the Full and SSE spectra (statistics 51.39 kg·y). The best fit to the SSE spectrum including 8 levels gave C≈1.3 events/kev.

But, in the whole, results of the PSA, given in [2], were disappointing (Table 1). The table presented the ratios of the numbers of SSE events to the numbers of events in the full spectrum for the 51.39 kg·y data. The numbers were calculated by a direct summation over the 5 kev intervals in the proper spectra from the [2]. The centers of the intervals corresponded to the energies of the peaks given in the table 7 of the [2]. One extra peak (2044 kev) was added, as its existence was well established . The line ∑min included points with number of events n≤6 and was used for the estimation of the SSE/FULL ratio for the constant (slow varying ) component of the background.

Table 1. Results of PSA: SSE/FULL ratios.

| E, kev | Peak, kev | ΔE, kev | ∑SSE, events | ∑FULL, events | sse/full |
|---|---|---|---|---|---|
| 2008-12 | 2010.7 | 5 | 18.0 | 58.0 | 0.31±0.07 |
| 2014-18 | 2016.7 | 5 | 11.0 | 58.8 | 0.19±0.06 |
| 2020-24 | 2021.8 | 5 | 17.2 | 48.1 | 0.36±0.09 |
| 2028-32 | ~2030 | 5 | 18.9 | 44.7 | 0.42±0.10 |
| **2036-40** | **2038.1** | **5** | **20.6** | **48.9** | **0.42±0.09** |
| 2042-46 | ~2044 | 5 | 12.0 | 30.0 | 0.40±0.11 |
| 2051-55 | 2052.9 | 5 | 23.1 | 48.4 | 0.48±0.10 |
| ∑min* | | 21 | 30.6 | 104.9 | 0.29±0.06 |

*) The ratio SSE/FULL=0.29 and $C_{SSE}$=1.3 events/kev were used to estimate constant components of the Full ($C_{FULL}$=4.5 events/kev) and the Rejected spectra.

Gamma-peaks contained only ~5% of the SSE [12], and it has been expected, that the SSE/FULL ratios for the peaks would be much less then the experimental values, 0.3-0.5 .

The two factors have increased the measured SSE/FULL ratios. First, the full gamma-spectrum contained a slowly varying Compton component, which included about (20-30)% of events due to the single scatterings of gamma-quanta. Shapes of such signals should be identical to the shapes of the $2\beta 0\nu$-decay signals and they were used for an education of the SSE identification programs [10,11]. The corrected SSE/FULL ratio for gamma peaks should be ~0.19 if one took into account this admixture. This value was achieved for the only one gamma-line (2016.7 kev), which was produced by a coincidence of two gamma-quanta [13]. An erroneous identification of a part of signals as the SSE ones was the second reason of the distortion.

If the 2038 kev peak were due to the 2β0ν-decay, the corrected for the Compton components value of the SSE/FULL ratio should be ~0.68. An efficiency of the SSE selection program was high [14] , and consequently the SSE/FULL ratio for this peak would be noticeably higher then the observed 0.42, which was identical to those values for the gamma-peaks . As no difference was found between the SSE/FULL ratios for the 2038 kev peak and it's neighbors, the results of PSA suggested the background origin of the 2038 keV line.

However there was a brief information in the same papers [2,3] which pointed out the unusual features of the peak. There was found empirically a subclass of events of the SSE spectrum (let us name it as the SSE Special Selection –SSE SS), when all the other lines were strongly suppressed, except the 2038 kev peak (fig. 3).

The details of the procedure have not been presented. Nevertheless the authors of [4] claimed the discovered empirical effect as a final proof for the observation of the $2\beta 0\nu$-decay of $^{76}$Ge with some higher value $T_{1/2}(0\nu) = (2.23\pm^{0.44}_{0.31})\cdot 10^{25}$ years.

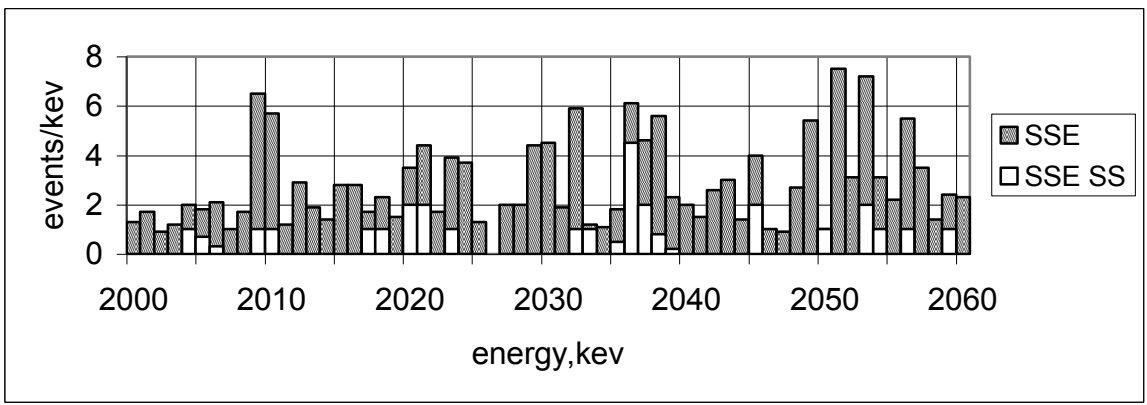

Fig.3. SSE and SSE SS spectra. The constant component was absent in the SSE SS spectrum.

It was difficult to estimate the validity of the above result as no details of forming the SSE SS spectrum were given. But in any case, to accept the claim one should explain the failure of the conservative PSA first of all (table 1).The neuronal net method [11] of the SSE selection was very effective [14]. Therefore the failure of the PSA meant an existence of some factor, which has not been taken into account.

PSA failure. The unresolved background levels.

The most simple and logistic explanation of the PSA failure would be the existence of an unknown background peak with energy close to the energy of $^{76}$Ge 2β0ν-decay. To check up this version, the spectrum of rejected with the PSA procedure events was reconstructed. It was calculated as the difference between the FULL and the SSE spectra of [3], N(rejected)=N(full)-N(SSE). A comparison of the REJECTED and SSE spectra in the energy interval around the 2038 kev peak indicated possible presence of the two different levels: at 2037.5 kev in the SSE spectrum and 2039.5 kev in the REJECTED spectrum (fig.4,5).

The result given in the table 1 for the 2038 kev peak (SSE/FULL=0.42) was calculated by an integration over the energy interval (2037-2041) kev which included partially both the above peaks. The same ratios calculated separately over the 3-kev intervals were $(SSE/FULL)_{2037}=0.68$ and $(SSE/FULL)_{2040}=0.32$. If one took into account the constant components of backgrounds, C=~4,2 events/kev for the FULL and C=~1,3 events/kev for the SSE spectra, the ratios for the peaks were $(SSE/FULL)_{2039}=0.96$ and $(SSE/FULL)_{2040}=0.10$. The first one was in agreement with the 2β0ν-decay origin of the 2037.5 kev peak. The second value pointed out that the peak at 2039.5 kev was due to a coincidence of two gammas, analogous to the origin of 2017.6 kev line.

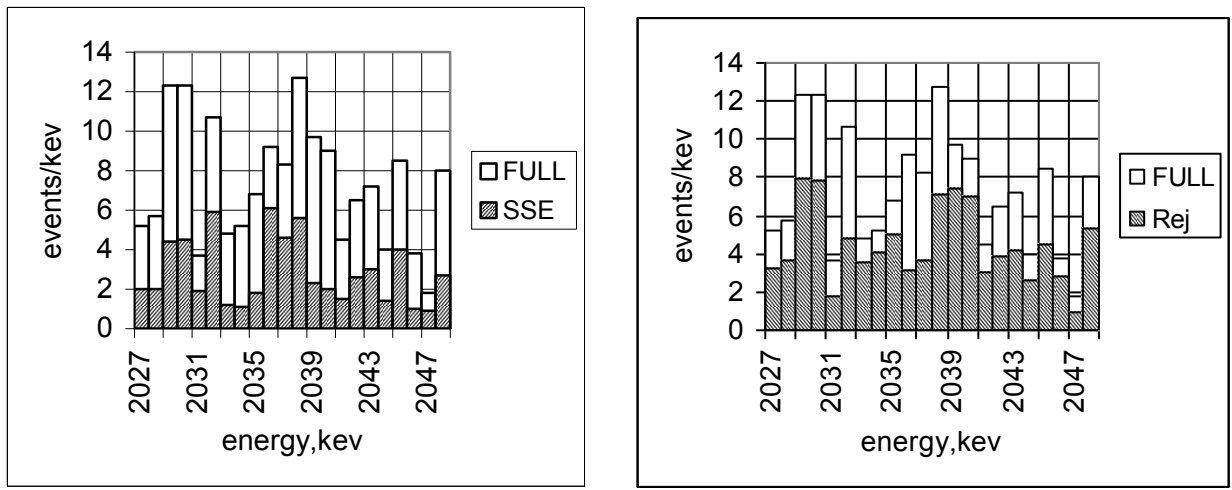

Fig.4,5. Comparison of the SSE and Rejected spectra in the energy interval in the vicinity of the 2038 kev peak. The constant components were estimated as C≈1.3 events/kev for the SSE spectrum, C≈3.2 events/kev for the Rejected one.

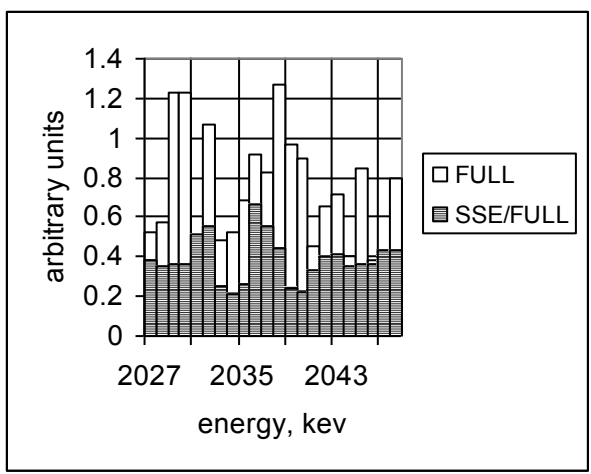

To get an additional information, the ratios $N_{SSE}/N_{FULL}$ with 1 kev binning were calculated for the energies around the original 2038 kev peak, (2027-2048)kev (fig.6).

Points near the peak demonstrated rather complicated energy structure. A minimum at ~2039 kev testified that the proper peak had the coincident nature. On the contrary the 2037.5 kev peak had a maximum SSE/FULL ratio, which was very close to the expected for the 2β0ν-decay level. But the existence of the second close in energy minimum at ~2034.5 kev was quite unexpected as it indicated presence of the second coincident line.

Fig.6. SSE/FULL ratios near the 2038 kev peak.

It was difficult to suppose that the 2038 kev peak in the full spectrum included three unresolved overlapping lines, two of which were unobserved earlier combined peaks ! To find such peaks in independent measurements with Ge detector would be a strong support of the above consideration.

Background peaks

Authors of [1-4] carefully analyzed possible radioactive sources of background, which could emit gamma-quanta with energies close to the energy of the $2\beta 0v$-decay of $^{76}$Ge. Such gamma-quanta have not been found, and it was the stabling block of their claim. It has not contradicted the above conclusion as both the indicated peaks seemed to be the combined ones. To find the combined peaks would be possible only with the special measurements, because a source should be placed just near the detector. The results of such measurements with $^{214}$Bi were given in [13].

The 2017 kev peak in the background spectra [2,3] was attributed to the 1408 kev+609 kev cascade coincidence. Still other combined peaks were not found.

Much better chances to observe such peaks presented the measurements with low-background Ge devices which contained an active anticoincidence NaI shielding [5-7]. Such configuration of the set-up provided an effective additional suppression of the gamma-background except the combined peaks: no gamma-quanta escaped the detector in this case.

The goal of the measurements [5] was also a search of the $2\beta 0v$-decay of $^{76}$Ge. The devices had usual heavy passive shields, and, beyond it, the detectors were surrounded with the NaI(Tl) anticoincidence counters. A comparison of the results of these measurements with the FULL spectrum indicated presence of an excess of events above background in the (2035-2040) kev region, which could be attributed to the more than one unknown peaks in the close vicinity to the energy of the $2\beta 0v$-decay of $^{76}$Ge (Fig.7)

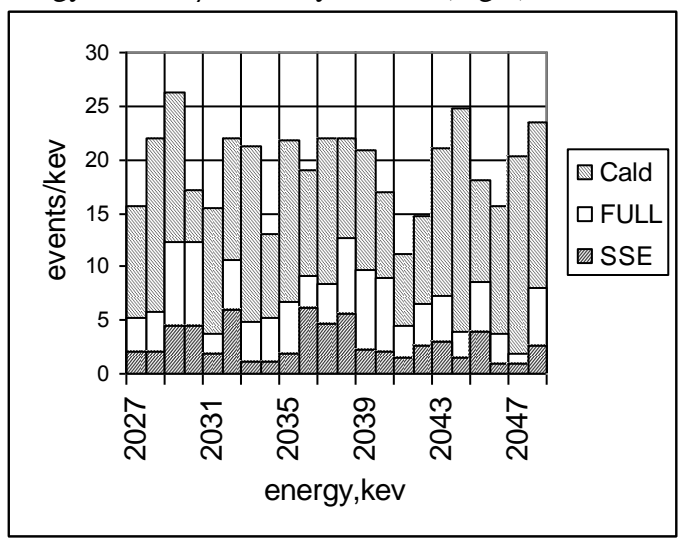

Fig.7. Caldwell, FULL, SSE spectra.

More definite results were given by the GEMMA measurements [7]. The GEMMA set up was a low-background Ge device for a search for the neutrino magnetic moment. It had a usual heavy passive shield, and the detector was placed inside it in a NaI(Tl) well. The set-up was operated at the Kalinin power station..

The device had extremely low background at the very low-energies which was necessary for the experiment. But index of the background at 2 Mev was rather high due to the operation near the reactor and not deep underground (~40 mwe): 50 times more that those of [2,3]. One could expect an observation of a set of combined peaks due to double coincidences of gamma-quanta in the detectors under these conditions.

An analysis of the GEMMA background at high energies was performed. The spectrum was collected within 9700 hours (~1.7 kg·y) (fig.8-11).

The figures demonstrated a part of GEMMA background at the energies close to the energy of the 2038.5 kev peak. The "bump" in the Caldwell spectrum was partially resolved onto the two peaks (fig8,9) just at the energies which had been predicted by the separation of the SSE and Rejected spectra (fig.4,5). The 2040 kev peak coincided with the maximums in the Rejected spectrum (fig.9) and the 2037.5 kev line was just in the middle between the two backgrounds peaks (fig.11). Positions of both the peaks in the GEMMA spectrum were also the same as positions of the two minimums, E = (2034-2036) kev and E = (2040-2042) kev, in the energy distribution of the N(SSE)/N(FULL) ratios presented in fig.6.

Thus the hypothesis has been supported that the 2038 kev peak in the full spectrum [2,3] was produced by overlapping of three unresolved peaks: ~2034.5 kev, ~2037.5 kev and ~2039.5 kev. Two of them (2034.5 kev and 2039.5 kev) were due to the double-coincidences of gamma-quanta in the detectors. The levels were too close together and could not be resolved in the full spectrum. Separation of the SSE and the Rejected spectra provided such possibility.

The improved result of the PSA for the 2037.5 kev line was in agreement with the 2β0ν-decay origin of it.

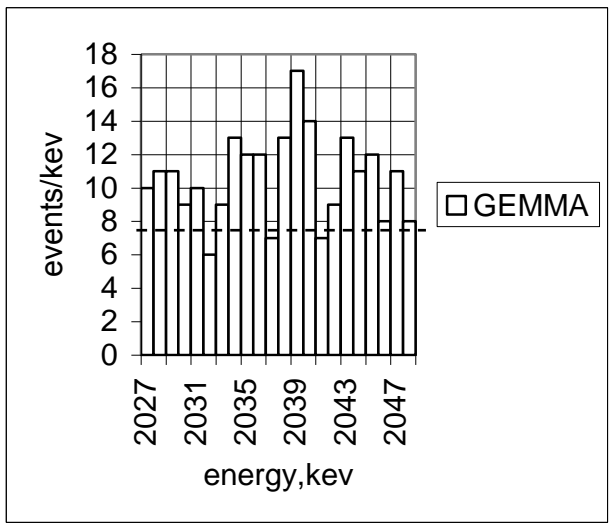

Fig.8. GEMMA spectrum. The dashed line shows the constant component.

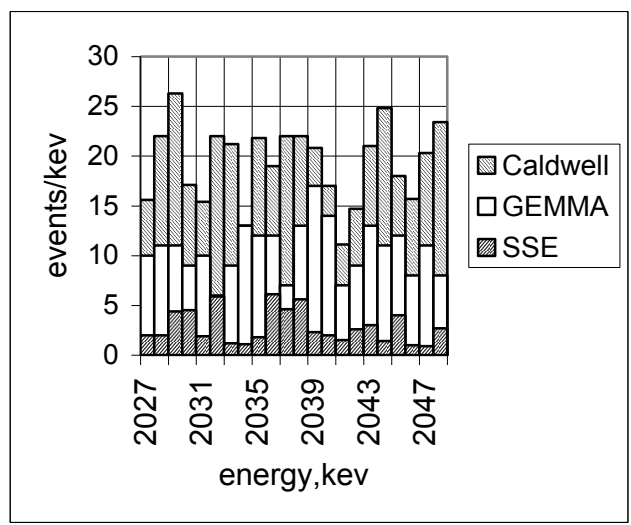

Fig.9. Caldwell, GEMMA, SSE spectra

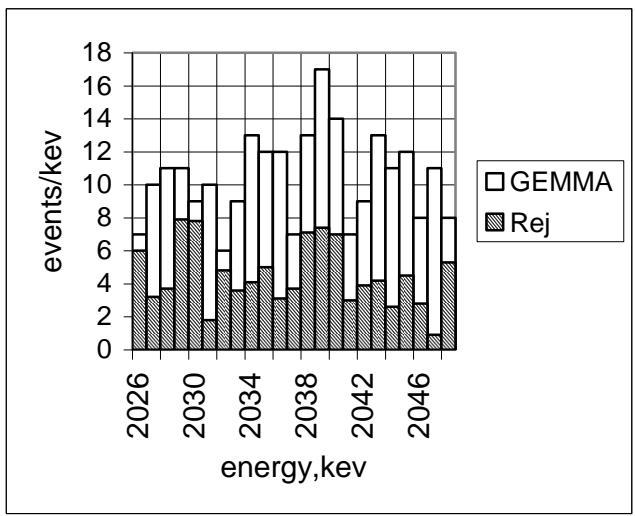

Fig.10. GEMMA and Rejected spectra.

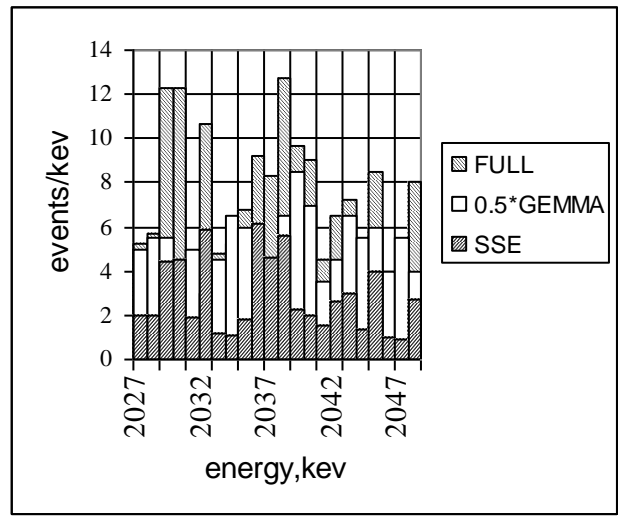

Fig.11. FULL, GEMMA, SSE spectra.

The SSE SS spectrum

The SSE SS spectrum was the only indication of the unusual origin of the 2038 kev peak (not from the gamma-background). Therefore the procedure of forming the spectrum need a special consideration.

There was a sequence of questions connected with the procedure, namely :
   a) a drastic reduction of background compare to the SSE spectrum, including disappearance of

the slow varying component (fig.3)

b) a loss of the essential part of events in the 2038.5 kev peak: 19.6 events in the FULL spectrum, 12.4 events in the SSE one,  7 events in the SSE spectrum

c) a shift of the peak position to the lower energies: 2038.5 kev – in the FULL spectrum, 2037.5 kev – in the SSE one, 2036.5 – in the SSE SS spectrum. The explanation through a "ballistic effect"[2,3,4] was wrong, as there was no any radial selection of signals (table 2)

Table 2. Radial distribution of the events [4].

| radius | ≤ 22 mm | > 22 mm |
|---|---|---|
| SSE | 11 events | 19 events |
| SSE SS | 2 events | 6 events |

The most difficult for an explanation phenomenon was the drastic reduction of the background in the SSE SS compare to the SSE spectrum (more than 7 times). Quite surprising was disappearance of the slow varying (constant) component.

This component contained events from the single Compton scattering. All the SSE selection programs were educated just through this process. The constant component in the SSE spectrum was evaluated as 1.3 events/kev (table 1). It was close to the expected minimum, and further decreasing of it should be connected with an uncontrolled abortion of the 2β-decay events. More than seven times suppression of the constant component by the program of SSE Special Selection indicated the inadequate data treatment.

The lack of information prevented the quantitative consideration of the SSE SS program. It was only clear from [4] that the SSE SS spectrum was a product of some combination of the sub-programs of the SSE selection . So only the qualitative conclusions could be made.

One could suppose that this combination produced a subtraction from the SSE spectra the gamma-background signals which were erroneously identified as the SSE events. To check up such possibility, an imitated spectrum was constructed which reproduced the possible procedure.

The SSE spectrum could be presented as SSE=FULL-Rejected. It was found empirically that the combination Imit=FULL-1.23*Rejected ≡ SSE – 0.23*Rejected  (fig.12) provided  the full suppression of the constant component of the background and the coincident gamma-peaks (the coefficient m=1.23 provided <Imit>=0 in the 5 kev energy interval around the 2017 kev peak). The imitation did not improve the statistics of the SSE data, but the most probable number of events in the 2037.5 kev peak could be estimated with the better accuracy, $N=14\pm3.7$ events ($N=12.4\pm3.7$ events [2,3]).

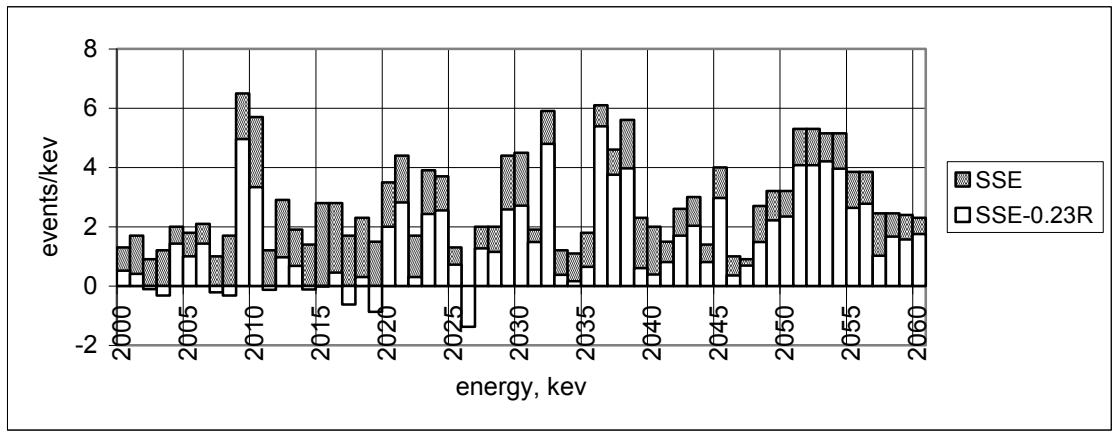

Fig.12. Combination Im=FULL-1.23*Rej provided the full suppression of the constant component and the coincident gamma-peaks.

By an analogy, the combination Imit=FULL-1.62*Rejected was chosen to provide the full suppression of the normal peaks (table 1) in the energy interval (2027-2048) kev (this energy interval of the SSE SS spectrum did not contain practically events except the 2038.1 kev line, fig.13). The result of the imitation was presented in fig.14.

The total number of events in the SSE SS peak and their energy distribution were fully reproduced. Almost a half of the events, which have been selected by the neuronal net program (fig.3,13), was aborted. A position of the peak shifted to 2036.5 kev.

Appearance of the two groups of negative numbers at 2034.4 kev and 2039.5 kev was expected as the imitation coefficient m=1.62 provided the full suppression of the SSE events from the normal gamma-peaks . The 2034 kev and 2040 kev peaks were supposed to be the combined ones. The suppression of the gamma-background near these peaks was overestimated as a result. It seemed that the SSE SS procedure of [2,3] was similar to the investigated imitation, and the resulting SSE SS spectrum contained essential systematic errors.

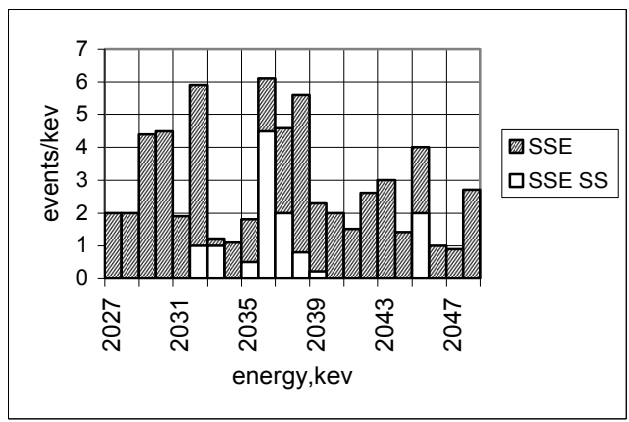 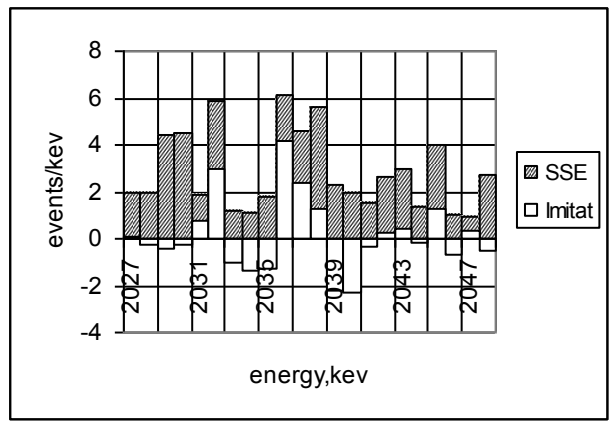

Fig.13. SSE and SSE SS spectra.   Fig.14. SSE and Im=FULL-1.62*Rej spectra

Summary

It was found that the mysterious 2038 kev line in the $^{76}$Ge background spectrum [2,3] was produced by overlapping of three unresolved peaks: ~2034.1 kev, ~2037.6 kev and ~2039.6 kev. Two of them (2034.1 kev and 2039.6 kev) were due to the double-coincidences of gamma-quanta in the detectors. The 2037.5 kev line could be a signal of the expected 2β0ν-decay transition of $^{76}$Ge according to the improved results of the pulse-shape analysis. Position of the line, 2037.56±0.56(stat)±1.2(syst) kev [4], was slightly below the expected, $E_{2\beta}$=2039.0±0.05 kev [8], but still within a limit for the poor statistics. The number of events in the peak, which was extracted from the 51.39 kg·y data with PSA, N=12.4±3.7 events, with the corresponding value of the lifetime for the 2β0ν-decay of $^{76}$Ge, $T_{1/2}$=1.98·10$^{25}$ years [2,3]. It might be only slightly corrected for an overestimation of the SSE background. The corrected value was N=14±3.7 events and the lifetime $T_{1/2}$=1.8·10$^{25}$ years .

Acknowledgement:
The author thanks Dr. A.S.Starostin for supplying him with the high-energy part of the GEMMA spectrum. An analysis of these data gave the crucial arguments for the unresolved overlapping peaks hypothesis.